\documentclass[twocolumn,showpacs,preprintnumbers,amsmath,amssymb,pre]{revtex4}

\newcommand{\BE}{\begin{equation}}
\newcommand{\EE}{\end{equation}}
\newcommand{\BA}{\begin{eqnarray}}
\newcommand{\EA}{\end{eqnarray}}

\usepackage{graphicx}% Include figure files
\usepackage{dcolumn}% Align table columns on decimal point
\usepackage{bm}% bold math
\preprint{Cond-mat/0211466}

\begin{document}

\title{Theoretical analysis of the focusing of acoustic waves by
two-dimensional sonic crystals}

\author{Bikash C. Gupta}
\author{Zhen Ye}\email{zhen@phy.ncu.edu.tw} \affiliation{Wave Phenomena
Laboratory, Department of Physics, National Central University,
Chungli, Taiwan 32054, R. O. C.}

%\date{February 2, 2002}
\date{\today}

\begin{abstract}

Motivated by a recent experiment on acoustic lenses, we perform
numerical calculations based on a multiple scattering technique to
investigate the focusing of acoustic waves with sonic crystals
formed by rigid cylinders in air. The focusing effects for
crystals of various shapes are examined. The dependance of the
focusing length on the filling factor is also studied. It is
observed that both the shape and filling factor play a crucial
role in controlling the focusing. Furthermore, the robustness of
the focusing against disorders is studied. The results show that
the sensitivity of the focusing behavior depends on the strength
of positional disorders. The theoretical results compare favorably
with the experimental observations, reported by Cervera, et al.
(Phys. Rev. Lett. 88, 023902 (2002)).

\end{abstract}

\pacs{43.20.+g, 43.58+z, 43.90.+v} \maketitle

\section{Introduction}

Photonic crystals \cite{review,rev2} (PC) are made of periodically
modulated dielectric materials, and most sonic crystals
\cite{sigal} (SC) are made up of materials with periodic variation
of material compositions. Both the photonic and sonic crystals
have been studied both intensively and extensively. It has been
suggested that PCs may be useful for various applications such as
antennas \cite{brown}, optical filters \cite{chen}, lasers
\cite{evans}, prisms \cite{lin1}, high-Q resonant cavities
\cite{lin2}, wave-guides \cite{kraus1}, mirrors \cite{kraus2} and
second harmonic generations \cite{mart}. And SCs could be used as
sound shields and acoustic filters
\cite{sanchez1,sanchez2,ye,ye1,rubio,caba1,rober,sanchis,kuswa}.
These applications mostly rely on the existence of photonic and
sonic bandgaps, and a majority of them is not concerned with the
linear dispersion region well below the first gap. Thus most of
earlier studies were focused on the formation of bandgaps and the
propagation inhibition of waves.

Recently, the interest in the low frequency region, where the
dispersion relation is linear, has just started. Since the
wavelength in this region is very large compared to the lattice
constant, the wave sees the media as if it were homogeneous, in
analogy with wave propagation in normal media. Consequently, it
has been proposed \cite{halevi} that the optical lenses could be
developed by use of PCs. However, no physical realization has been
made so far. Along the same line of the thought, it was suggested
that SCs may also be used to build acoustic lenses to converge the
acoustic waves. A necessary condition to be satisfied for
constructing an acoustic lens is that the acoustic impedance
contrast between the SC and the air should not be large; otherwise
acoustic waves will be mostly reflected. Once this condition is
satisfied, the converging lens can be either convex or concave
depending on whether the sound speed in the SC is smaller or
greater than that in the air. Recently, it has been experimentally
found \cite{cervera} that the sound speed in two dimensional
regular arrays of rigid cylinders in air is less than that in air
and the impedance contrast between the structure and air is not
large. The authors presented the physical realization of two
refractive devices, namely, a Febry-Perot interferometer and an
acoustic convergent lens.

In this paper we carry out numerical simulation on the focusing of
acoustic waves by SCs. We wish to theoretically analyze the
experimental results reported in \cite{cervera}. Since the
multiple scattering technique has been successfully applied
earlier \cite{ye} to reproduce some experimental results on
acoustic propagation and scattering in SCs, we will use this
technique to study in detail about the focusing effect of acoustic
waves by sonic crystals. There are several parameters associated
with the SCs, such as the shapes and the arrangements of the
lattice, the acoustic frequencies, and the filling factors etc. We
will show how the focusing is controlled by these parameters. In
addition, the sensitivity of focusing effects to positional
disorders will also be studied. The paper is organized as follows.
The formalism is presented in Sec. II, followed by the results and
discussions in Sec. III. Here we will reproduce theoretically the
experimental observations in \cite{cervera}. A summary concludes
the paper in Sec. IV.

\section{Formalism}

Consider $N$ straight cylinders located at $\vec{r_i}$ with $i=1,
2, ...N$ to form either a completely random or regular array. An
acoustic line source transmitting monochromatic waves is placed at
$\vec{r_s}$. The scattered wave from each cylinder is responsible
to the total incident wave composed of the direct wave from the
source and the multiply scattered waves from other cylinders. The
final wave reaching a receiver located at $\vec{r_r}$ is the sum
of the direct wave from the source and the scattered waves from
all the cylinders. Such a scattering could be solved exactly,
following Twersky \cite{twersky}. The essential procedure is
presented below and the detail could be found in Ref.\cite{ye1}.

The scattered wave from the $j$th cylinder can be written as \BE
p_s(\vec{r}, \vec{r_j}) = \sum_{n=-\infty}^{\infty} i \pi A_n^j
H_n^{(1)}(k|\vec{r}-\vec{r_j}|) e^{in\phi_{\vec{r}-\vec{r_j}}},
\label{eq:1} \EE where $k$ is the wave number of the medium,
$H_n^{(1)}$ is the $n$th-order Hankel function of first kind, and
$\phi_{\vec{r}-\vec{r_j}}$ is the azimuthal angle of the vector
$\vec{r}-\vec{r_j}$ relative to the positive $x$ axis.

The total incident wave around the $i$th scatterer,
$p_{in}^i(\vec{r})$, which is a superposition of the direct
incident wave from the source and the scattered waves from all
other scatterers, can be expressed in terms of the Bessel function
(of first kind) as

\BE p_{in}^i(\vec{r}) = \sum_{n=-\infty}^{\infty} B_n^i
J_n(k|\vec{r}-\vec{r_i}|) e^{in\phi_{\vec{r}-\vec{r_i}}}.
\label{eq:3} \EE Using the addition theorem for Bessel function,
the scattered waves $p_s(\vec r, \vec{r_j})$ for each $j \ne i$
can be expressed in terms of the modes with  respect to the $i$th
scatterer and thus one leads to \BE B_n^i=S_n^i + \sum_{j=1,j \ne
i}^{N} C_{n}^{j,i} \label{eq:10} \EE with \BE S_l^i = i \pi
H_{-l}^{(1)}(k|-i l\phi_{\vec{r_i}}|) \label{eq:9} \EE and \BE
C_n^{j,i} = \sum_{l=-\infty}^{\infty} i \pi A_l^j
H_{l-n}^{(1)}(k|\vec{r_i}-\vec{r_j}|) {\rm
exp}[i(l-n)\phi_{\vec{r_i}-\vec{r_j}}]. \label{eq:7} \EE Matching
the usual boundary conditions for each cylinders, one obtains: \BE
B_n^i = i \pi \Gamma_n^i A_n^i \label{eq:15} \EE where
$\Gamma_n^i$ are transfer matrix relating the acoustic properties
of the scatterers and the media (see Ref. \cite{ye1} for the
expression). All coefficients $A_n^i$ may easily be obtained by
solving the Eq.~(\ref{eq:10}) along with Eq.~(\ref{eq:15}) and
hence the total wave at any desired point outside the cylinders
may  be obtained as \BE p(\vec{r}) = p_0(\vec{r}) + \sum_{i=1}^{N}
\sum_{n=-\infty}^{\infty} i \pi A_n^i H_n^{(1)} (k|\vec{r} -
\vec{r_i}|) e^{i n \phi_{\vec{r}-\vec{r_i}}}. \label{eq:20} \EE
The acoustic intensity (field) is the square module of the
transmitted wave where the normalized transmission is given as $T
\equiv p/p_0$. It is worth to mention that the total wave
expressed by Eq.~(\ref{eq:20}) incorporates all orders of multiple
scattering and the formalism is valid for both regular and random
arrangements of the scatterers.

\section{Results and discussions}

In the following computation, we consider regular arrays of rigid
cylinders in air. The system we use is from and has been detailed
in \cite{cervera}. There are $N$ uniform rigid cylinders of radius
$a$. The area occupied by the cylinders per unit area is $\beta$.
Therefore, the lattice constant can be expressed as $d=
\sqrt{(2\pi a^2/\sqrt 3 \beta )}$ for a triangular lattice
arrangement of the cylinders. For the corresponding random
arrangement of the cylinders, $d$ represents the average distance
between the nearest neighbor cylinders. Unless otherwise
mentioned, the lengths will be scaled by $d$ and the frequency
will be scaled as $ka$, to make them dimensionless. We will mainly
consider the triangular lattice and use the parameters from
\cite{cervera} for the filling factor, and the radii of the
cylinders.

\input{epsf}
\begin{figure}[hbt]
\epsfxsize=2.5in\epsffile{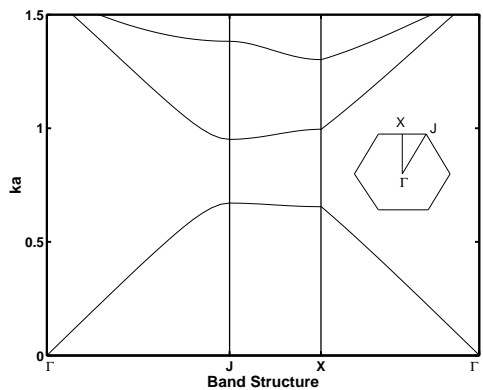} \caption{\label{fig1} \small
Band structure for the two dimensional triangular lattice of the
rigid cylinders in air.}
\end{figure}

Figure~\ref{fig1} reproduces the band structure for the two
dimensional triangular lattice from \cite{cervera}. The filling
fraction is $0.2024$, the radius of each cylinder is $1.5 cm$, and
thus the lattice constant is $\approx 6.35 cm$. We see that the
dispersion in the low frequency region is linear for both the
directions. Our calculations will be confined in the linear low
frequency region only.

\begin{figure}[hbt]
\epsfxsize=2.75in\epsffile{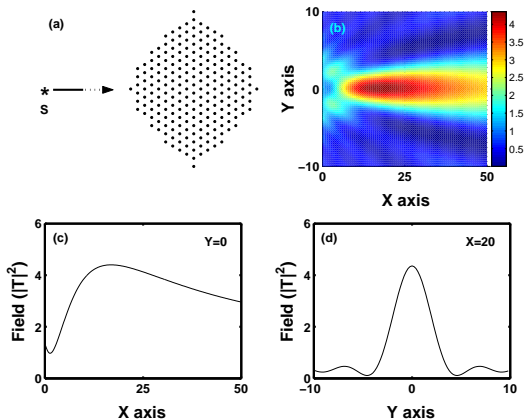} \caption{\label{fig2} \small
(a) The line source denoted by {\bf S} and the arrangement of
cylinders; (b) two dimensional spatial distribution of the
transmitted intensity ($|T|^2$) on the right side of the crystal;
(c) the variation of the intensity in (b) along the $x$ axis at
$y$=0 and (d) the variation of the intensity in (b) along the $y$
axis at $x$=20.}
\end{figure}

In accordance with the experimental setup described in
\cite{cervera}, the conceptual setup of the source and the lattice
arrangement of the cylinders for the computation are shown in
Fig.~\ref{fig2}(a). The line source is denoted by {\bf S}. Here
the triangular arrangement in the experiment is taken, and the
lenticular shape is identical to that used in the experiment
\cite{cervera}. The lattice constant for the lattice is $\sim 6.35
cm$, the same as considered in the experiment. The source is
placed at $100 d$ away from the left side of the lattice; this
distance is far enough to ensure that the incident wave on the
lattice behaves as a plane wave. The wave with frequency of
$ka=0.486$ ($\sim 1700 Hz$) is incident on the triangular lattice
of lenticular shape from the left, and the transmitted intensity
($|T|^2$) is calculated on the right side of the lattice. The two
dimensional spatial distribution of the transmitted intensity is
shown in Fig.~\ref{fig2}(b). Here the $x$ axis is a horizontal
line towards right, and the $y$ axis is placed vertically upward.
The origin is at a distance twice the radius of the cylinders to
the right most cylinder, so to avoid the diverge when plotting the
spatial distribution. The same arrangement of the coordinates is
also used in the following figures, except for Fig.~\ref{fig7}. We
see that the focusing of the transmitted wave is evident, and is
in agreement with the experimental observation \cite{cervera}.
Since the maximum intensity point, which we assume as the focal
point, is not clear from Fig.~\ref{fig2}(b), we have computed the
variation of the field intensity along the $x$ axis with $y=0$,
and also along the $y$ axis with $x=20$. The results are shown in
Figs.~\ref{fig2}(c) and \ref{fig2}(d). Fig.~\ref{fig2}(c) shows
that there is a peak at $x \sim 14$. The intensity variation along
the $y$ axis shows a significant peak at $y=0$. Here we observe
that the intensity is better confined along the $y$ axis than
along the $x$ axis. The theoretical results are in a good
agreement with the experimental observation.

\input{epsf}

\begin{figure}[hbt]
\epsfxsize=2.75in\epsffile{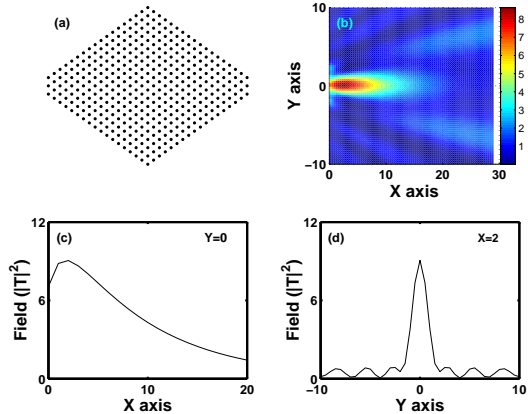} \caption{\label{fig3}
\small(a) The arrangement of the cylinders; (b) the two
dimensional spatial distribution of the transmitted intensity
($|T|^2$) on the right side of the lens; (c) the variation of the
intensity in (b) along the $x$ axis at $y$=0 and (d) the variation
of the intensity in (b) along the $y$ axis at $x$=2.}
\end{figure}

In view that our numerical calculation can reproduce the
experimentally observed focusing effect of sound, we are
encouraged to further explore possible focusing effects with
different shapes of SCs. Here we consider the lenticular shape
shown in Fig.~\ref{fig3}(a). The size of the lens along the $y$
direction is same as that for Fig.~\ref{fig2}, whereas the size
along the $x$ axis is larger. Thus, the edges of the lattice are
much smoother than that in the above lens. Here the source is
again kept at $100d$ away from the left side of the lens, and
other parameters are the same as used above. The transmitted
intensity calculated on the right side of the lens is shown in
Fig.~\ref{fig3}(b). We see that the focusing is apparently better
than that in the above SC of lenticular shape. We also plot the
intensity variation along both the $x$ axis with $y=0$ and the $y$
axis with $x=2$. The results are shown in Figs.~\ref{fig3}(c) and
\ref{fig3}(d) respectively. Fig.~\ref{fig3}(c) clearly shows that
the intensity is peaked at about $x=2$ and $y=0$, then decreases
more rapidly compared to that in Fig.~\ref{fig2}(c).
Fig.~\ref{fig3}(d) also shows a narrower peak when compared with
Fig.~\ref{fig2}(d). In this case, the focal length is shorter and
the intensity at the focal point is much stronger than those in
the above case. Again, the intensity region is more extended along
the $x$ axis compared to that along the $y$ axis.

\input{epsf}

\begin{figure}[hbt]
\epsfxsize=2.75in\epsffile{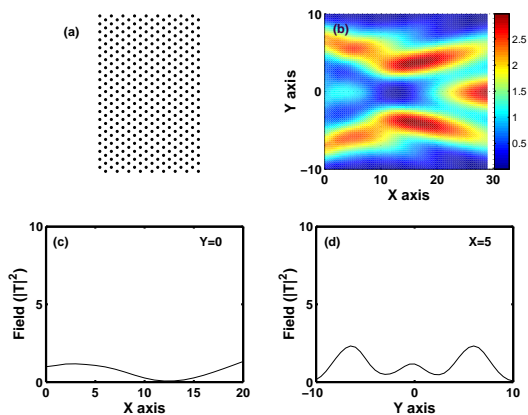} \caption{\label{fig4} \small
(a) The arrangement of the cylinders; (b) the two dimensional
spatial distribution of the transmitted intensity ($|T|^2$) on the
right side of the crystal; (c) the variation of the intensity in
(b) along the $x$ axis at $y=0$, and (d) the variation of the
intensity in (b) along the $y$ axis at $x=5$.}
\end{figure}

For comparison, we have also studied the transmission through a
rectangular lens shown in Fig.~\ref{fig4}(a). The results are
depicted in Fig.~\ref{fig4}(b). In this case, no unique focusing
is observed. The transmitted intensity is more or less spreading
all over the $xy$ plane. This can be better seen from
Figs.~\ref{fig4}(c) and \ref{fig4}(d), where the intensity
variations are plotted along the $x$ axis and $y$ axis
respectively. These results also agree with the experimental
observation \cite{cervera}.

\input{epsf}

We also performed the similar calculations for the lattices with
the regular square arrangement of the cylinders and found that the
results are similar.

\begin{figure}[hbt]
\epsfxsize=2.75in\epsffile{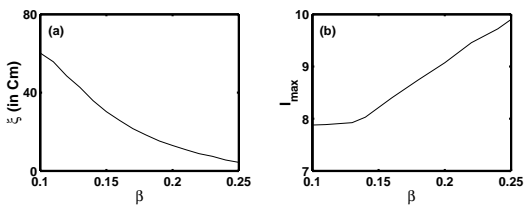} \caption{\label{fig5} \small
(a) The variation of the focal length as a function of filling
fraction ($\beta$); and (b) the variation of  Maximum intensity,
${\rm I}_{{\rm max}}$ (the intensity at the focal point) as a
function of the filling fraction.}
\end{figure}

Figure \ref{fig5}(a) shows the variation of the focal length,
i.~e. the distance between the focal point and the center of the
lattice, as a function of the filling factor. We see that the
focal length ($\xi$ in cm) decreases rapidly with the increasing
filling factor ($\beta$) up to $\beta=0.2$, then tends to saturate
for higher filling factors. This decrease of the focal length with
the filling factor is reasonable as the the effective sound speed
in the SC decreases with increase of the filling factor, to be
discussed later. The variation of the intensity at the focal point
(${\rm I}_{{\rm max}}$) as a function of filling factor is also
shown in Fig.~\ref{fig5}(b). Therefore, in addition to the shapes,
the filling factor is another parameter that effectively controls
the focal point and focusing intensity.

\input{epsf}

From the above study on lattices of different shapes, we conclude
that lattice shape is very crucial to obtain the focusing effect.
The quality of focusing and the focal length could be adjusted by
varying the shape of the crystal or the filling fraction.

\begin{figure}[hbt]
\epsfxsize=2.75in\epsffile{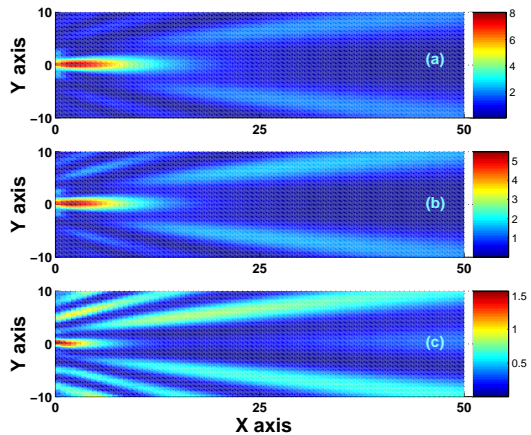} \caption{\label{fig6} \small
Two dimensional spatial distribution of the transmitted intensity
($|T|^2$) on the right side of the lattice with (a) weak, (b)
moderate, and (c) strong positional disorders. The shape of the
lattice and the filling factor used here are the same as that
depicted for Fig.~\ref{fig3}.}
\end{figure}

In practice, it is useful to know the robustness of the focusing
against possible disorders in the system. Here we consider the
positional disorder of the cylinders. Since the wavelength is
larger than the lattice constant in the present cases, one might
intuitively conclude that the positional disorder has no effects.
This is untrue. We have performed the calculation for three cases:
the weak, moderate, and strong positional disorders. For the weak
and moderate disorders, every cylinder can randomly take a
position within a circle with radius of 1.4 and 15 percent of the
lattice constant about its regular position, while for the strong
disorder, the cylinders may take positions randomly within the
boundary of the regular crystal as long as no two cylinders
overlap with each other, i.~e. completely random.
Figs.~\ref{fig6}(a) and (b) show the two dimensional spatial
distribution of the transmitted intensities for the weak and
moderate disorders. It is clear that the focusing effect is hardly
affected by the presence of weak disorder. Fig.~\ref{fig6}(c)
shows the distribution of the transmitted field for the strong
disorder case. Here the the intensity is highly suppressed all
over the $xy$ plane and also the intensity at various places in
the $xy$ plane becomes comparable to that at the maximum intensity
region. Therefore, we conclude that slight or mild deviations from
perfectly regular arrangements when designing acoustic lenses will
not affect the focusing drastically; therefore it will be easier
to build the acoustic lens. However, the strong disorder does
affect the focusing significantly. It is interesting to note that
although the focusing nature is greatly reduced in the presence of
strong disorders, the spatial pattern of the scattered intensity
remains almost unchanged. This is because at the long wavelength
region, the waves will not be able to discern the detailed
structure of the crystal, and therefore the scattering pattern is
mainly controlled by the shape of the crystal. The reduction in
the transmitted intensity indicates that the waves are reflected
backward. This is confirmed below.

\input{epsf}

To understand such drastic difference for the focusing between the
regular lattice and the disordered lattice we plot the spatial
distribution of back scattered intensities for the regular and the
disordered lattices and they are shown in Fig.~\ref{fig7}(a) and
\ref{fig7}(b) respectively. Here the $x$ axis is a horizontal line
towards right, and the $y$ axis is placed vertically upward.  The
origin is at a distance twice the radius of the cylinders to the
most left cylinder. We observe that the distribution is symmetric
for the regular case, but asymmetric for the disordered case.
Compared to the forward scattering pattern for the disordered case
in Fig.~\ref{fig6}, the backscattering is more sensitive to the
random configuration. This is understandable since when waves are
backscattered, the travelling path will be twice that when going
forward, thereby causing more interference between scattered
waves. The intensity for the disordered case appears to be
stronger than in the regular case. The quantitative difference is
further illustrated by Figs.~\ref{fig7}(c) and \ref{fig7}(d) where
the variations of back scattered intensities for both the regular
and disorder cases are plotted along the $x$ axis with $y=0$ and
the $y$ axis with $x$=0 respectively. While the solid curve
represents the variation of back scattered intensities for the
regular case, the dashed curve for the disordered case. Here, it
is clear that the intensity strength for the disorder case is much
higher. In fact, the ratio of the maximum back scattered intensity
between the two cases can be as high as 15. In other words, in the
disordered configuration, the transmission is highly prohibited,
in agreement with the earlier observation\cite{ye}. In this case,
the lattice does not appear to be transparent to acoustic waves.

\input{epsf}
\begin{figure}[hbt]
\epsfxsize=2.75in\epsffile{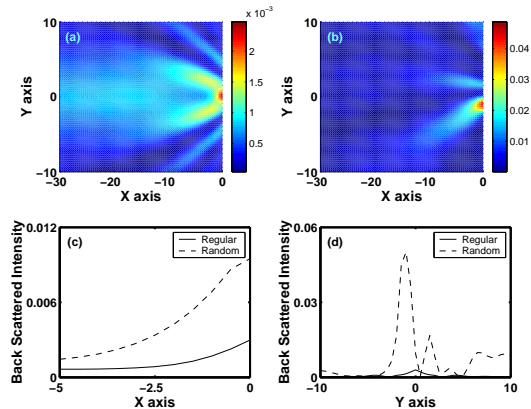} \caption{\label{fig7} \small
Two dimensional spatial distributions of the back scattered
intensities from (a) the regular lattice, and (b) one completely
random configuration of the lattice; no configuration averaging is
taken. The shape of the crystals for both cases is the same as
that in Fig.~\ref{fig3}.}
\end{figure}

\begin{figure}[hbt]
\epsfxsize=2in\epsffile{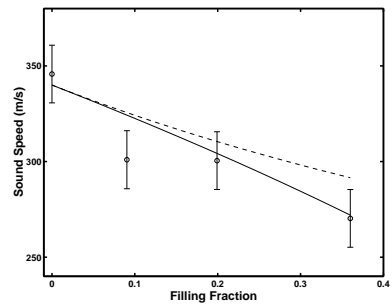} \caption{\label{fig8} \small The
variation of sound speed in the SC with the filling fraction: The
circles with error bars represent the experimental values, the
dashed curve represents the theoretical estimate in
\cite{cervera}, and the solid curve represents our estimate.}
\end{figure}

Finally we note on the acoustic refraction from the sonic crystal
formed by the rigid cylinders in air. The focusing effect shown
above is essentially, as already argued by the authors in
\cite{cervera}, due to the reduction of the sound velocity within
the SC. A simple model has been proposed by the authors to explain
this reduction. Upon examination, we found that the results from
the model agrees with the experimental data rather well in the low
filling regime. A deviation between the theoretical and
experimental data is obviously seen, referring to Fig.~2 in
\cite{cervera}, for high filling factors. Here we would like to
propose a modified model to account for the experimental data. At
the frequency considered in the experiment\cite{cervera} and the
simulation above, the wavelength is much bigger than the lattice
constant. Therefore the sonic crystal may be regarded as an
effective air-filled medium, i.~e. the air around each cylinder is
regarded as an effective air which takes into account of the
presence of other cylinders. The presence of the rigid cylinders
is actually to reduce the occupation of the air per unite area
without the cylinders, and is replaced by the effective air
medium. This scheme may be called the self-consistent procedure.
The effective mass density can be then estimated from \BE \rho_{e}
= \rho_{air} + \beta \rho_{e},\EE where $\rho_{air}$ and
$\rho_{e}$ are the densities of the air and the effective medium
respectively. Thus the density of the effective medium is solved
as \BE \rho_{e} = \frac{\rho_{air}}{1-\beta}.\EE For a given
pressure, the sound speed in an air filled medium is proportional
to the inverse of the square root of the density, i.~e. $c\sim
\sqrt{\frac{1}{\rho}}$, thus we obtain the sound speed of the
sonic crystal \BE c_{e} = c_{air}\sqrt{1-\beta}. \label{eq:c}\EE
Obviously, by $1-\beta \approx 1/(1+\beta)$ for small $\beta$, the
result reduces to that in \cite{cervera} for small filling
factors. In Fig.~\ref{fig8}, we show the comparison among the
experimental data, and the theoretical results from the present
model and that in\cite{cervera}. Here we clearly see that present
modified model reproduces the experimental data remarkably well,
better than the original model.

\input{epsf}

\section{Summary}

Through rigorous calculations we have successfully reproduced the
focusing of acoustic waves by the sonic crystal made of rigid
cylinders in air, observed by the recent experiment. We have
examined the focusing effect for crystals of different shapes. It
has been shown that the shape of the crystal plays a crucial role
on the quality of focusing. The dependance of the focal length on
the filling factor is also examined. It is found that the focusing
becomes brighter and the focal length decreases as the filling
factor increases. The robustness of the focusing against the
presence of disorders is also investigated. The results show that
the robustness of the focusing behavior depends on the strength of
disorders. Finally, a model describing the sound speed in the SC
made from rigid cylinders arranged periodically in air as a
function of filling fraction is proposed. An excellent agreement
is found when the results from the model are compared with the
experimental data.

\acknowledgments{This work received support from National Science
Council of Republic of China (Grant No: NSC 90-2811-M008-004).}

\end{document}